\documentclass[
 aps,
 pre,
 amsmath,amssymb,
 reprint,
showkeys
]{revtex4-2}

\usepackage[T1]{fontenc}
\usepackage[utf8]{inputenc}
\usepackage{amsmath}
\usepackage{amssymb}
\usepackage{graphicx}
\usepackage{refstyle}
\usepackage{multirow}
\usepackage[
  citecolor=blue,
  colorlinks,
  linkcolor=blue,
  urlcolor=blue,
]{hyperref}
\usepackage[dvipsnames]{xcolor}
\usepackage{dcolumn}
\usepackage{bm}
\usepackage{soul}
\usepackage[T1]{fontenc}
\usepackage{xcolor}

\begin{document}
\title{Game-environment feedback dynamics in growing population: Effect of finite carrying capacity}

\author{Joy Das Bairagya}
\email{joydas@iitk.ac.in}
\affiliation{
	Department of Physics,
	Indian Institute of Technology, Kanpur 208016, India
}
\author{Samrat Sohel Mondal}
\email{samrat@iitk.ac.in (Corresponding author)}
\affiliation{
  Department of Physics,
	Indian Institute of Technology, Kanpur 208016, India
}
\author{Debashish Chowdhury}
\email{debch@iitk.ac.in}
\affiliation{
 Department of Physics,
	Indian Institute of Technology, Kanpur 208016, India
}
\author{Sagar Chakraborty}
\email{sagarc@iitk.ac.in}
\affiliation{
  Department of Physics,
	Indian Institute of Technology, Kanpur 208016, India
}

\date{\today}
	\begin{abstract}
	The tragedy of the commons (TOC) is an unfortunate situation where a shared resource is exhausted due to uncontrolled exploitation by the selfish individuals of a population. Recently, the paradigmatic replicator equation has been used in conjunction with a phenomenological equation for the state of the shared resource to gain insight into the influence of the games on the TOC. The replicator equation, by construction, models a fixed infinite population undergoing microevolution. Thus, it is unable to capture any effect of the population growth and the carrying capacity of the population although the TOC is expected to be dependent on the size of the population. Therefore, in this paper, we present a mathematical framework that incorporates the density dependent payoffs and the logistic growth of the population in the eco-evolutionary dynamics modelling the game-resource feedback. We \textcolor{black}{discover a bistability in the dynamics: a finite carrying capacity can either avert or cause the TOC depending on the initial states of the resource and the initial fraction of cooperators.} In fact, depending on the type of strategic game-theoretic interaction, a finite carrying capacity can either avert or cause the TOC when it is exactly the opposite for the corresponding case with infinite carrying capacity.	
	\end{abstract}
	\maketitle
	\section{Introduction}
	\label{sec:introduction}
 
Resources	are finite; selfishness is infinite. Consequently, selfish rational individuals---who defect from helping the others with a motive of maximizing their own utilities---cause losses to the cooperators in the population and lead to over-exploitation of the shared common resources and, thus, bring forth the unfortunate tragedy of the commons (TOC)~ \cite{malthus, lloyd1833two,hardin1968commons}. Uncontrolled population growth~\cite{hardin1968commons}, water pollution and water crisis~\cite{shiklomanov2000appraisal}, pollution of the earth's atmosphere~\cite{jacobson2002atmospheric}, property rights, communal rights or state regulation~\cite{ostrom1999coping}, and wildlife crimes~\cite{pires2011preventing} are a few of the examples of the TOC. The TOC leads a system to a degraded state of the shared resource such that in the long run, the evolutionary as well as the ecological fitnesses of all the individuals are adversely affected~\cite{bellard2012impacts}. 

Moreover, the TOC is not restricted to merely human society; it can be witnessed in evolutionary systems where the concept of rationality is arguably non-existent: For example, intra- and inter-plant competitions for root proliferation give rise to a competition for the nutrition and water intake from soil by the plants~\cite{gersani2001tragedy,zea2006tragedy,o2008games}; also, the TOC is witnessed in microbes vying for nutrients~\cite{2003BC,MacLean2008,SMITH2019R442,10.1371/journal.pone.0186119} or aerial oxygen~\cite{rainey2003evolution}. Another intriguing example is witnessed in the colonies of eusocial Hymenoptera where the conflict among the immature females trying to develop into a queen in order to gain greater direct reproductive fitness imposes a productivity cost on the colony by reducing the common resource---the workforce, i.e., the number of workers~\cite{bourke1999kin,wenseleers2003caste,wenseleers2004worker,wenseleers2004tragedy,hughes2008social}.

Any depleted resource around a population has adverse effect on the population size if the growth of the population depends on it; or in slightly technical terms, if the carrying capacity~\cite{murray2011mathematical,verhulst1838,verhulst1845,Verhulst1977} of the population is tied to the resources under consideration. The carrying capacity refers to the upper limit of the size of the population that can be sustained. It is easy to envisage that this feedback between the resources and the ecology of the population---in the light of the finite carrying capacity---can give rise to interesting eco-evolutionary dynamics that is suitably modelled through (evolutionary) game-theoretic ideas~\cite{replica,smith1982book,nowak2006evolutionary,hofbauer1998book} describing the strategic interactions between the cooperators and the defectors.  This paper focuses on this aspect of the TOC. The resources present around a population constitutes the environment for the population; hence, in line with the existing terminology in the literature, we use resources and environment synonymously. In other words, this paper is concerned with the game-environment feedback in a growing population while keeping in mind that the carrying capacity is practically always finite. 

The state of the environment---i.e., how replete or depleted it is---can change the preferences of the individuals (henceforth, to be called players in accordance with the game-theoretic terminology): As the environment degrades, the cooperation tendency must increase in the population in order to avert the TOC. In fact, enforcing cooperation in various ways is an obvious and well-studied mechanism for averting the TOC	~\cite{hardin1968commons,cox1985no,wade1985common}, even in chimpanzees~\cite{koomen2018chimpanzees}. If one considers the players to be von Neumann--Morgenstern rational~\cite{von1953theory}, the change of preferences can be quantitatively showcased through their utilities or the payoff matrices corresponding to the interaction between the players. Interesting recent works~\cite{weitz2016oscillating, lin2019prl, tilman2020evolutionary} have explicitly mathematized this idea to elaborately study the fate of the commons in the resulting eco-evolutionary dynamics.

In the context of the growing population, however, the concept of the environment needs to be scrutinized little bit more closely. Specifically, one should appreciate that there are two somewhat similar aspects of the environment: The environment may be seen as a combination of two resources---one, ecological resource, that is directly responsible for the growth of the population size and the other, common shared resource, that is under threat from the overexploiting defectors. These two resources are not necessarily mutually exclusive. However, we find that, for our purpose, the case where the two resources are mutually exclusive is qualitatively similar to the other cases.

\textcolor{black}{ The conceptually non-trivial formalism that we adopt in this paper is mathematically minimal while keeping all the essential aforementioned ideas incorporated into it. \textcolor{black}{ Specifically}, the evolutionary dynamics is taken to be governed by the paradigmatic replicator dynamics~\cite{taylor1978mb,schuster1983jtb,schuster1985bbpc,hofbauer1998book,page2002jtb,cressman2003book}, the population growth is considered logistic~\cite{murray2011mathematical,verhulst1838,verhulst1845,Verhulst1977}, and the state of the common shared resource is also considered to be essentially logistic in nature. In the next section, we elaborately discuss the deterministic mathematical model on which our investigation of this paper rides.}

\section{The model}	
\textcolor{black}{ Let there be a set of $\mu$ distinct strategies that can be adopted by any individual member of the consumer population. 
Let $N_i(t)$ be the number, and $x_i(t) =N_i/N$ ($1 \leq i \leq \mu$) denote the corresponding fraction, of the consumer population at time $t$ that adopts the $i$-th strategy. Alternatively, $x_i$ can also be interpreted as the probability or frequency of the $i$-th strategy being used. The vector ${\boldsymbol x}(t) \equiv (x_1(t), x_2(t), \cdots, x_{\mu}(t))$ defines the state of the consumer population. The total population of the consumers at time $t$ is $N(t) = \sum_{i} N_{i}(t)$ that defines the size of the population. Let ${\boldsymbol n}(t) \equiv (n_1(t), n_2(t), \cdots, n_{\nu}(t))$ denotes the state of the $\nu$ distinct shared resource pools in the environment where, for convenience, we assume that variables $n_{j}$ are normalized so as to lie in the unit interval, i.e., $0 \leq n_{j}(t) \leq 1$.}

\textcolor{black}{ Thus, the state of the composite system, consisting of the consumer population and the shared resource pools is given by $\boldsymbol{\sigma}(t)\equiv (N(t), {\boldsymbol x}(t),{\boldsymbol n}(t))$. The dynamics of the system is governed by $\mu-1$ differential equations for ${\boldsymbol x}(t)$ (since $\sum_{i=1}^{\mu} x_i = 1)$, $\nu$ differential equations for ${\boldsymbol n}(t)$ and a single equation for $N(t)$.}

\textcolor{black}{ Having succinctly presented the general formulation of the systems of our interest, we now turn to the specific setting that is conducive to studying the tragedy of the commons. \textcolor{black}{ More specifically, in the simplest nontrivial setting, the rise of the defectors who lead to the TOC can be exemplified through the prisoner's dilemma game ($T > R > P > S$)~\cite{1965_RC}. The prisoner's dilemma game is a one-shot two-player--two-strategy game ($\mu=2$) in which defecting is the only symmetric Nash equilibrium~\cite{nash1950pnas,wagner2013} that is non-Pareto-optimal~\cite{1896_Pareto}; mutual cooperation could fetch the players comparatively more reward.} Its normal bimatrix form is as follows:
 \begin{eqnarray*}  
\centering
\begin{tabular}{cc|c|c|}
		& \multicolumn{1}{c}{} & \multicolumn{2}{c}{{Player $2$}}\\
		& \multicolumn{1}{c}{} & \multicolumn{1}{c}{Cooperate} & \multicolumn{1}{c}{\,\,\,\,Defect\,\,\,\,}\\\cline{3-4} 
		\multirow{2}*{{Player $1$}} & Cooperate & $R,R$ & $S,T$ \\\cline{3-4}
		& Defect & $T,S$ & $P,P$ \\\cline{3-4} 
\end{tabular}\quad
\end{eqnarray*}
where the first and the second elements (real numbers) in each cell respectively are the payoffs of player $1$ and player $2$ and $R$, $S$, $T$, and $P$ respectively refer to Reward, Sucker's payoff, Temptation, and Punishment. Thus, here we identify $x_1$ and $x_2$ as the fractions of `{\it cooperators}' and `{\it defectors}', respectively. Furthermore, for simplicity, we choose $\nu=1$. Therefore {the set of three dynamical variables---${\boldsymbol x}(t) \equiv (x_1(t),x_2(t)=1-x_1(t))$, ${\boldsymbol n}(t) \equiv n(t)$, and $N(t)$---together describe the state $\boldsymbol\sigma(t)$ of the composite system}. In fact, since $x_1$ and $x_2$ are not independent variables, we henceforth use $x(t)=x_1(t)$ as the only variable to specify the state of the population, wherever convenient.}

\textcolor{black}{ We now explain the motivation for the choice of the actual effective payoff matrix used in our model in a step by step manner. In the first step, let us begin with the simplest situation where $R$, $S$, $T$, and $P$ are the payoffs realized ignoring the environmental resources and the size of the population. In this case the payoff matrix, ${\sf U}({n},N)$, for a focal player would have the $n$- and $N$-independent form
\begin{eqnarray}
{\sf U}({ n}={\rm constant}, N={\rm constant})={\sf U}=
  \left[ {\begin{array}{cc}
   R & S \\
   T & P \\
  \end{array} } \right], 
\label{eq-U}
\end{eqnarray}
where the elements are time-independent numbers. {In the next step}, in order to extend the payoff matrix to include the effect of the state of the environment, we introduce an effective payoff matrix $\tilde{{\sf U}}(n)$ that has the form 
\begin{eqnarray}
\tilde{\sf{U}}(n) = (1-n) {\sf U_0} + n {\sf U_1}.  
\end{eqnarray} 
Here ${\sf U_k}$ ($k\in\{0,1\}$) is the shorthand notation for ${\sf U}({ n}={k}, N=K\rightarrow\infty)$. In more explicit words, in the presence of fixed infinite population, the payoff matrix reduces to ${\sf U}_{0}$ in the limit of $n=0$, which corresponds to poor environmental resource; whereas in the opposite limit $n=1$  of rich resource it reduces to ${\sf U}_{1}$. {Guided by the form (\ref{eq-U})}, a natural parametrization of ${\sf U}_{0}$ and 
${\sf U}_{1}$ is 
\begin{eqnarray}
   {\sf U}_{0}=
  \left[ {\begin{array}{cc}
   R_0 & S_0 \\
   T_0 & P_0 \\
  \end{array} } \right] 
\label{eq-U0}
\end{eqnarray}
and 
\begin{eqnarray}
   {\sf U}_{1}=
  \left[ {\begin{array}{cc}
   R_1 & S_1 \\
   T_1 & P_1\\
  \end{array} } \right].
\label{eq-U1}
\end{eqnarray}} 

\begin{figure}
		\centering
		\includegraphics[width=1.0\linewidth]{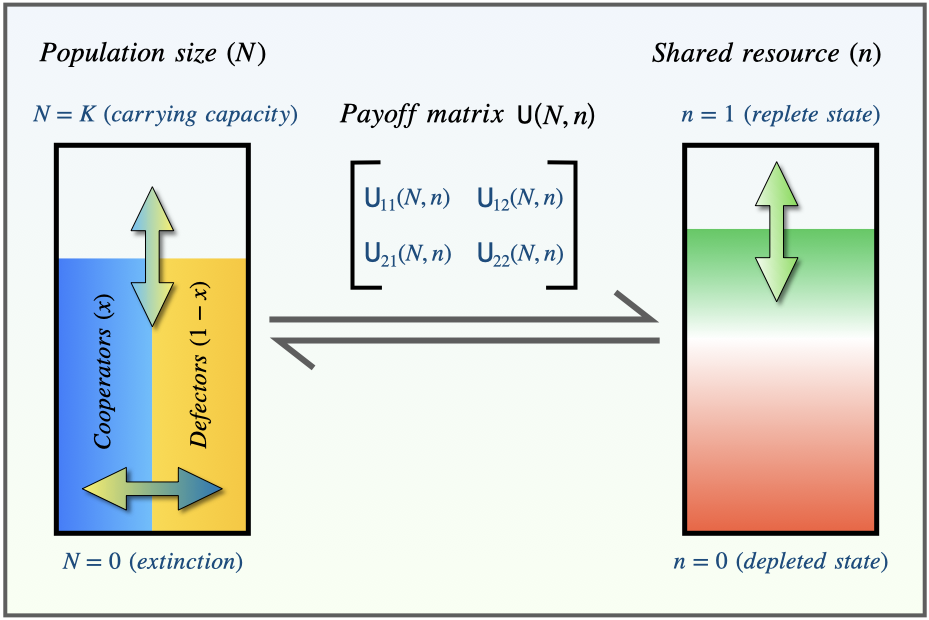}
		\caption{Schematic diagram of the environmental feedback via games: A finite population (left box) of size $N$ with carrying capacity $K$, has $x$ fraction of the cooperators (blue) and $1-x$ fraction of the defector (yellow). It affects the shared resource (right box) $n$ whose state moves between richer (greener) as well as poorer (redder) states via the game-dynamics---governed by the payoff matrix $\textsf{U}(N,n)$---between the cooperators and the defectors.}	
		\label{fig:fig1}
	\end{figure}

\textcolor{black}{ The new key element of our model is the {\it finite carrying capacity} $K$ whose effect is incorporated next in the third step of our generalization of the payoff matrix. We assume that the finite carrying capacity further modifies the payoff matrix to the following form:
\begin{eqnarray}
{\sf U}(n,N) = (1-n) \biggl[\biggl(1-\frac{N}{K}\biggr){\sf U}_{0} \biggr] + n \biggl[\biggl(1-\frac{N}{K}\biggr){\sf U}_{1} \biggr]. \quad
\label{eq-PayOffModel}
\end{eqnarray}
We note that the matrix ${\sf U}(n,N)$ is independent of ${\boldsymbol x}$ because we are considering matrix games~\cite{cressman2014pnas} where ${\boldsymbol x}$-dependence {enters through} the fitness $f_i$---an individual consumer's {\it fitness} upon adopting the $i$-th strategy is given by
\begin{equation}
f_{i}(\boldsymbol\sigma) = \sum_{j=1}^{2} {\sf U}_{ij}({ n},N) x_{j};\quad i=1,2.
\end{equation}
Since the fitter population is expected to grow at a faster rate, it is quite natural to assume 
\begin{eqnarray}
	\frac{dN_i}{dt}=f_{i}(\boldsymbol\sigma(t))N_i.
	\label{frequency}
\end{eqnarray}
Summing Eq.~(\ref{frequency}) over all the strategies, we get	
\begin{eqnarray}
\frac{dN}{dt} &=&  N\sum_{i=1}^{2} \sum_{j=1}^{2} {\sf U}_{ij}(n,N) x_{i}x_j,
\label{eq-dNdt}
\end{eqnarray}
where $x_i$ is to be recognized as $N_i/N$. Making use of  Eq.~(\ref{frequency}) and Eq.~(\ref{eq-dNdt}), the equation for the evolution of $x_i$ is then simply
\begin{eqnarray}
\frac{dx}{dt} &=&  x\left[\sum_{j=1}^{2} {\sf U}_{1j}(n,N) x_{j} - \sum_{i=1}^{2}\sum_{j=1}^{2} {\sf U}_{ij}(n,N)  x_ix_{j}\right].\qquad
\label{eq-xEqModel1}
\end{eqnarray}
which is the replicator equation. Here $\sum_{j=1}^{\mu} f_{j}x_j$ is the mean fitness of the population.} \textcolor{black}{  Thus, the population evolves under replication-selection process modelled appropriately by the time-continuous replicator equation.} 

\textcolor{black}{  While the replicator equation can be derived in various equivalent ways~\cite{rice1961book,hofbauer1998book,page2002jtb,traulsen2005prl,2021MCChaos}, the continuous replicator equation was originally introduced in the evolutionary game theory to model the frequency dependent selection. In order to model density dependent selection~\cite{novak2013density} into it, we have incorporated the finite carrying capacity explicitly. It quantifies the implicit ecological resource. In addition, the dependence of the rate of evolution of the replicators on the state of an additional common shared resource must be modelled separately.}

\textcolor{black}{ To this end, we need to provide a dynamics for the variable ${n}(t)$. In this model, for simplicity, we ignore the any intrinsic dynamics of the resource population in the absence of the consumers. Instead, only the consumption (or enhancement) of the resource by the consumer population is assumed to alter the resource population. In this scenario, the cooperators help in augmenting the resource, while the defectors act to degrade it. If $\theta>0$ is the ratio of this enhancement rate to the degradation rate, it has been shown~\cite{weitz2016oscillating} that a simple model for the state $n$ of the shared resource  may be written as,
	\begin{equation}
	\frac{dn}{dt}=\epsilon [\theta x - (1-x)]n(1-n),
	\label{environment}
	\end{equation}
where parameter $\epsilon \ll 1$ because the environment is assumed to change relatively slowly compared to the strategists' frequencies.}

\textcolor{black}{ Eqs.~(\ref{eq-dNdt})--(\ref{environment}),  together with Eq.~(\ref{eq-PayOffModel}) as the payoff matrix, constitute the set of dynamical equations that mathematically describes the eco-evolutionary dynamics of the composite system's state, $\boldsymbol{\sigma}(t)\equiv (N(t), { x}(t),{ n}(t))$, in the presence of finite ecological resource quantified by a finite carrying capacity $K$. FIG.~\ref{fig:fig1} presents the set-up schematically. }

\section{Results}	
The main aim of our study is to explore the dependence of TOC on the types of strategic interactions and initial conditions. 

\textcolor{black}{ Eq.~(\ref{eq-PayOffModel}) enforces the intuitive expectation that the state of the resource affects the way the players interact. Since we are interested in the TOC, we assume that in the fully replete case the players play the prisoner's dilemma game (i.e., ${\sf U_1}$ corresponds to the prisoner's dilemma) but as degradation starts the cooperation can, in principle, ensue. Essentially this means that the form of ${\sf U_0}$ could deviate from the prisoner's dilemma. In general, depending on how the preference of the players change as degradation of the shared resource takes place, $U_0$ could be the payoff matrix for any of the four classes of games~\cite{Rapoport1967,Hummert2014,Pandit2018,Hilbe2018,Mittal2020}. {The games are classified into four types} based on  the correspondence of the Nash equilibria with cooperation and defection:  {(i)} in the \emph{harmony game}, the Nash equilibrium is mutual cooperation; {(ii)} in the \emph{anti-coordination} game, there exists a unique mixed symmetric Nash equilibrium in which the players play a mixed strategy randomized over the pure strategies; {(iii)} in the \emph{prisoner's dilemma}, the mutual defection is the unique Nash equilibrium; and {(iv)} in the \emph{coordination game}, there are two symmetric pure Nash equilibria [(cooperate, cooperate) and (defect, defect)] and one mixed symmetric Nash equilibrium like the one in the anti-coordination games.}

While we want to fully investigate how the finite ecological resource affects the eco-evolutionary dynamics, a specific curiosity of ours is whether any counter-intuitive situation appears where finite $K$ leads to the TOC even though infinite $K$ averts the TOC. 

\subsection{\textcolor{black}{ Linear stability analysis}}

It is easy to see that if a same additive constant is added to all the elements, the replicator equation remains invariant. Hence, for the purpose of our goal, we can restrict ourselves to only positive values of the payoff matrix elements. This means that $N$-component of phase velocity, $\dot{N}$, is always positive for $(0<N<K)$. Through this restriction, one can observe that there cannot be any periodic orbit in the phase space of the set of three eco-evolutionary dynamical equations because for that to occur it is required that generically all the components, $(\dot{x},~\dot{n},~\text{and}~\dot{N})$, of the phase velocity alter sign along such an orbit. The absence of the periodic orbits also mean the absence of unstable periodic orbits that must be densely embedded in a chaotic attractor for the later to exist. In other words, no chaotic attractor is also expected in the system of equations under consideration. These considerations naturally dictates us to exclusively confine our attention towards the fixed point solutions of the system.

In the model, there are three {\it non-isolated} sets of fixed points: 
\begin{itemize}
\item $(n^*=0,x^*\in[0,1],N^*=K)$: {completely depleted resource};
\item $(n^*=1,x^*\in[0,1],N^*=K)$: {completely replete resource};
\item $(n^*\in[0,1],x^*=1/(1+\theta),N^*=K)$: {partially replete resource}.
\end{itemize} 
So far as the linear stability of these non-isolated fixed points are concerned, we find the following:
\begin{itemize}
\item $(n^*=0,x^*\in[0,1/(1+\theta)),N^*=K)$: {stable};
\item $(n^*=0,x^*\in(1/(1+\theta),1],N^*=K)$: {unstable};
\item $(n^*=1,x^*\in[0,1/(1+\theta)),N^*=K)$: {unstable};
\item $(n^*=1,x^*\in(1/(1+\theta),1],N^*=K)$:  {stable};
\item $(n^*\in[0,1],x^*=1/(1+\theta),N^*=K)$: {stable or unstable, depending on the exact payoff matrix structure}.
\end{itemize} 
Other than these three set of non-isolated fixed points, there are some {\it isolated} unstable fixed points that in the format  $(n^*,x^*,N^*)$ are given by: $(0,0,0)$, $(0,1,0)$, $(1,1,0)$, $(0,P_0-S_0/(P_0+R_0-S_0-T_0),0)$, $(1,P_1-S_1/(P_1+R_1-S_1-T_1),0)$, and $([T_0-R_0+(P_0-S_0)\theta]/[T_0-R_0-T_1+R_1+(P_0-S_0-P_1+S_1)\theta],1/(1+\theta),0)$. All these fixed points are unstable and not of practical importance to our investigation.

\subsection{\textcolor{black}{ Relevant partition of the parameter space}}
Our next aim is to find out {\it how and when the stable fixed points} are physically attained as the system evolves with time. The full non-linear evolution can be studied only numerically. To this end, we numerically evolve the system for a wide range of initial conditions and parameter that we list shortly.

But first, following past works~\cite{weitz2016oscillating, tilman2020nc}, we introduce four parameters that can be intuitively interpreted as four distinct types of incentives for changes in strategies: 
\begin{subequations}
\begin{eqnarray}
\Delta^k_{RT} &\equiv& R_k-T_k,\\
\Delta^k_{SP} &\equiv& S_k-P_k,
\end{eqnarray}
\end{subequations}
where $k\in\{0,1\}$ specifies whether the payoff elements are for $n=0$ or $n=1$ case. In the literature~\cite{wang2015universal, tudge2015tale, arefin2020social,ito2018scaling},  $-\Delta^k_{RT}$ and $-\Delta^k_{SP}$ are known  as the gamble-intending dilemma strength and the risk averting dilemma strength respectively. We also introduce a parameter---relative dilemma strength, $\delta_k\equiv\Delta^k_{RT}/\Delta^k_{SP}=(-\Delta^k_{RT})/(-\Delta^k_{SP})$---that quantifies by what multiplicative factor a player has more affinity to defect against a cooperator than against a  defector.

As discussed earlier, we can envisage four  exhaustive, mutually exclusive classes listed below for which we present the results separately: 
\begin{itemize}
\item Harmony game: $R > T$ {and} $S > P$, i.e, $\Delta_{RT} >0$ {\and} $\Delta_{SP} > 0$.
\item Anti-coordination game: $R < T$ {and} $S > P$, i.e, $\Delta_{RT} < 0$ {and} $\Delta_{SP} > 0$.
\item Prisoner's~ dilemma: $T > R > P > S$, i.e, $\Delta_{RT} <0$ {and} $\Delta_{SP} < 0$.
\item Coordination~ game: {$R > T \geq P > S$, i.e, $\Delta_{RT} >0$ {and} $\Delta_{SP} < 0$}.
\end{itemize}
Since we want to analyse the dependence of TOC on the strategic interactions modelled by all possible types of ${\sf U_0}$, it is convenient to present the results on a plane spanned by $\Delta^{0}_{RT}=0$ and $\Delta^{0}_{SP}=0$. Thus the $\Delta^{0}_{RT}-\Delta^{0}_{SP}$ plane is divided into four quadrants identified by the four aforementioned different types of games, each having a distinct structure of the corresponding payoff matrix~\cite{Rapoport1967,Hummert2014,Pandit2018,Hilbe2018,Mittal2020}, {viz.}, that of the harmony game, the anti-coordination game, the prisoner's dilemma, and the coordination game.

Subsequently, drawing a line with slope $\delta_0 = \delta_1$ and another line with slope $-\delta_0= \theta$, the $\Delta^{0}_{RT}-\Delta^{0}_{SP}$ plane gets finally divided into seven distinct regions. In each of these regions thus obtained, we present the fate of any arbitrary initial condition $(x_0,n_0)$ for a finite value of carrying capacity and also for infinite carrying capacity~\cite{weitz2016oscillating}. Essentially, we find out whether an initial condition evolves to reach an attractor with $n=0$ (TOC) or with $n\ne0$ (averted TOC); the $x$-component of the attractor is the cooperator fraction in the final state.

\subsection{\textcolor{black}{ Numerical results}}
\begin{figure*}
		\centering
		\includegraphics[width=1.0\linewidth]{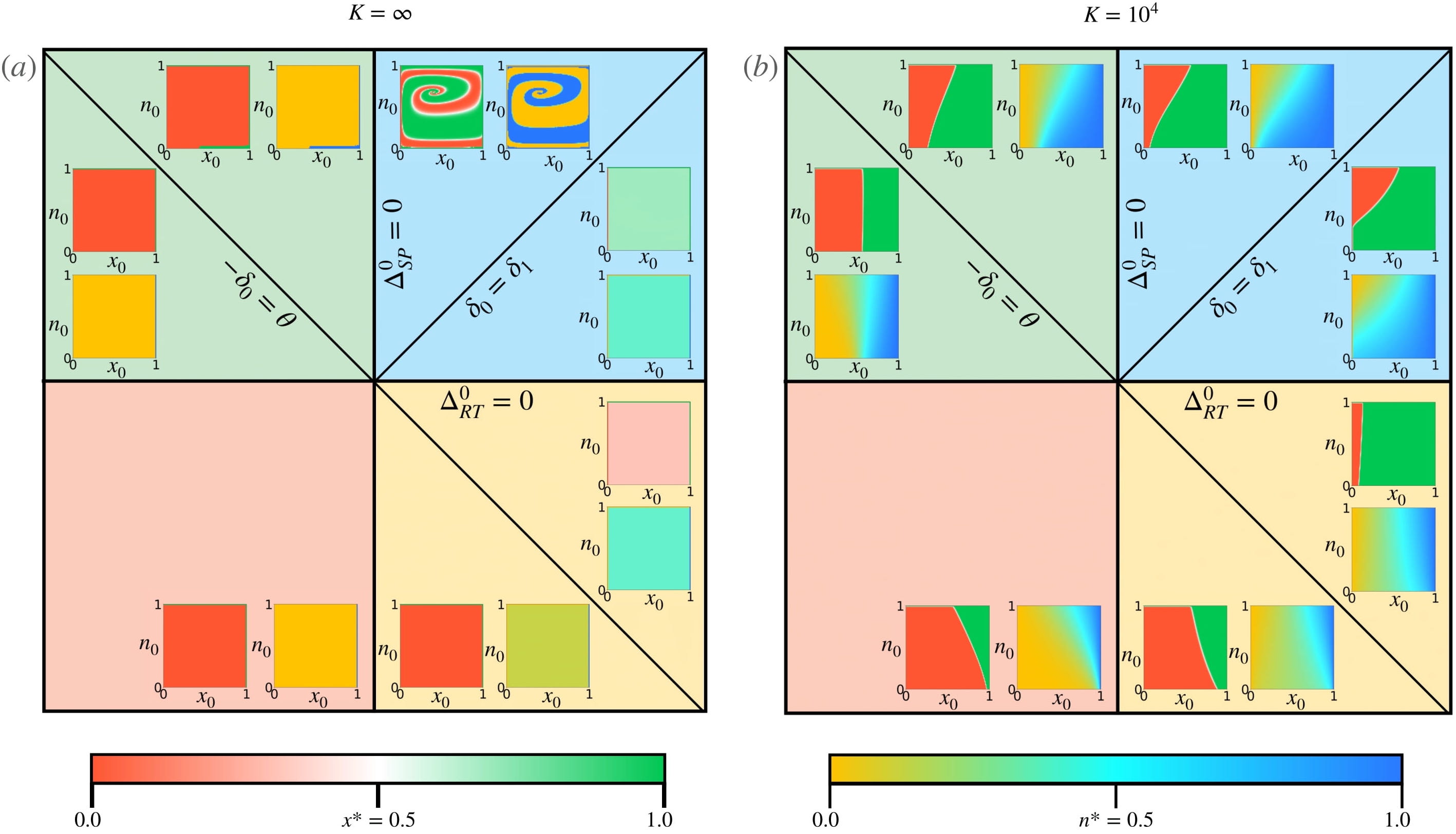}
		\caption{\textcolor{black}{ TOC is crucially dependent on the carrying capacity, the initial conditions, and the type of strategic interactions: Each of the two panels shows four regions---light blue, light yellow, light red and light green respectively corresponding to the matrix ${\sf U_0 }$ for the harmony game, the anti-coordination game, the prisoner's dilemma and the coordination game. The black lines demarcate seven different regions [for $K=\infty$ in subplot (a), and for $K=10^4$ in subplot (b)]. In each of these regions, using the common color codes shown below the panels, we present $x^*$ and $n^*$---the final cooperator fraction and the state of the shared resource respectively---realized for $100\times100$ uniformly distributed initial conditions $(x_0,n_0)$. The bluer the plot, the more is cooperator-fraction; the greener the plot, the more replete the shared resource is. Specifically, in the plots for $n^*$, red indicates realization of the TOC,  green indicates complete aversion of the TOC, and other colours indicate partial aversion of the TOC. In all cases with finite carrying capacity $K$, these plots for the final resource state, $n^*$, show bistability: Every initial condition either ends up at $n^*=0$ (TOC) or $n^*=1$ (averted TOC) as time evolves. We remark that the swirling red inside green in the plot for $n^*$ showcases the oscillatory-TOC~\cite{weitz2016oscillating}. Here we have fixed $N_0=10^2$  (other values give similar results).}}
		\label{ToC}
	\end{figure*}

In FIG.~\ref{ToC}, we pictorially summarize the comprehensive results thus obtained. While we discuss the details in the subsequent sections, one point is crystal clear from the figure: Unlike the case of the infinite carrying capacity, in the presence of the finite carrying capacity, the TOC is strongly dependent on the initial states of the population and the shared resource.

For generating the figures, we evolve the system by fixing\\
\\ 
${\sf U_1}=\tiny{\begin{bmatrix} 
			4 & 1 \\
			4.5 & 1.25\\
	\end{bmatrix}}$ (the prisoner's dilemma),\\ 
\\
$\theta =1.5$ and $\epsilon=0.1$; any change in these values do not qualitatively effect the results reported herein as is expected. For the sake of concreteness, we further chose following ${\sf U_0}$ in the above mentioned seven distinct regions (see FIG.~\ref{ToC} as well):
\begin{itemize}
\item For the harmony game with $\delta_0<\delta_1$: \\
${\sf U_0}=\tiny{\begin{bmatrix} 
3.5 & 1 \\
2 & 0.75\\
\end{bmatrix}}$.
\item For the harmony game with $\delta_0>\delta_1$:\\
 ${\sf U_0}=\tiny{\begin{bmatrix} 
3.5 & 1 \\
3 & 0.05\\
\end{bmatrix}}$. 
\item For the anti-Coordination game with $-\delta_0<\theta$:\\
 ${\sf U_0}=\tiny{\begin{bmatrix} 
3& 1 \\
3.5 & 0.05\\
\end{bmatrix}}$.
\item For the anti-Coordination game with $-\delta_0>\theta$:\\ 
${\sf U_0}=\tiny{\begin{bmatrix} 
3.5& 1 \\
2 & 1.25\\
\end{bmatrix}}$. 
\item For the prisoner's dilemma:\\ 
${\sf U_0}=\tiny{\begin{bmatrix} 
2& 0.05 \\
3.5 & 1\\
\end{bmatrix}}$. 
\item For the coordination game with $-\delta_0<\theta$:\\
${\sf U_0}=\tiny{\begin{bmatrix} 
3.5& 0.05\\
3 & 1\\
\end{bmatrix}}$. 
\item For the coordination game with $-\delta_0>\theta$:\\
 ${\sf U_0}=\tiny{\begin{bmatrix} 
3.5& 1.25 \\
2 & 1\\
\end{bmatrix}}$. 
\end{itemize}
We use the fourth order Runge--Kutta scheme for time-evolving the eco-evolutionary system. The system of differential equations evolve from $t=0$ to $t=1000$ in step-size of $dt=0.1$. We start with a population size of $N(t=0)=N_0=100$. So far as the choice of $x(t=0)=x_0$ and $n(t=0)=n_0$ are concerned, we choose $100\times100$ different initial conditions taken as the grid points of the uniformly spaced square grid of side $0.01$ units spanning the entire $x$-$n$ unit square. 
\subsubsection{\textcolor{black}{ $K\rightarrow\infty$: Pure effect of payoff structure}}

\textcolor{black}{ First, in FIG.~\ref{ToC}(a), we present the case of the infinite carrying capacity where only the payoff structure is {expected to decide the conditions under which} the TOC is averted. The corresponding results are exactly in line with what is known in the existing literature~\cite{weitz2016oscillating}, although with an interesting technical difference that we discuss in Sec.~\ref{sec:mech}. }

\textcolor{black}{ We present a short summary of the results for this special case: (i) The TOC is inevitable
when (a) ${\sf U}_0$ is that for the prisoner's dilemma or the coordination game; (b) ${\sf U}_0$ is for the anti-coordination game with $-\delta_0>\theta$. (ii) When ${\sf U}_0$ corresponds to the anti-coordination game with $-\delta_0<\theta$, the TOC is partially averted in the sense that an intermediate state of the resource---along with a mixed population of the cooperators and the defectors---is asymptotically arrived at. (iii) Most interesting is the case when ${\sf U_0}$ corresponds to the harmony game. The corresponding region in FIG.~\ref{ToC}(a) is divided by a line $\delta_0=\delta_1$; as $K\rightarrow\infty$,  there is an oscillatory TOC above this line whereas the TOC is partially averted below this line. Thus, ignoring the detailed features, we can conclude that in the case of the infinite carrying capacity ($K\rightarrow\infty$), the TOC is never completely averted (i.e., $n\ne 1$ asymptotically at all times) irrespective of the payoff structure.}
\subsubsection{\textcolor{black}{ Effect of finite carrying capacity}}
We find a finite carrying capacity is capable of averting the TOC which occurs in the corresponding counterpart with infinite carrying capacity: Specifically, this happens when ${\sf U_0}$ corresponds to the prisoner's dilemma, the coordination game, and the anti-coordination game with $-\delta_0>\theta$. One could intuitively argue that comparatively less number of individuals would be unable to exploit the shared resource fully and hence the TOC is possibly averted when the carrying capacity is finite. The case of the finite carrying capacity is presented in FIG.~\ref{ToC}(b).

More interesting are the cases where ${\sf U_0}$ corresponds to the harmony game and the anti-coordination game with $-\delta_0<\theta$: One finds that while in the case of the infinite carrying capacity the TOC is averted---at least partially ($0<n^*<1$) or periodically (oscillatory-TOC)---making the carrying capacity finite yield a set of initial conditions for which the TOC is, quite counter-intuitively, realized.

From Eq.~(\ref{eq-dNdt}) and Eq.~(\ref{environment}), we note $dn/dN\propto (1-N/K)^{-1}$. This implies that change in the state of the resource with the change in the size of population is more if the carrying capacity is finite. In other words, the individuals---both cooperators and defector---are capable of bringing more change in the state of the resource as they replicate. Consequently, if a finite population has a large fraction of cooperators (say, $x\rightarrow1$) to begin with, they may be able to positive change the state of the resource compared to the case of infinite population where the TOC is inescapable. While we explain in the next section how the interplay between the growth rates of $x$ and $n$ leads to this effect, it should be immediately clear that this argument is independent of the form of ${\sf U_0}$ as is validated in FIG.~\ref{ToC}(a). Intriguingly, it means that even if the prisoner's dilemma is played always (i.e., both ${\sf U_0}$ and ${\sf U_1}$ both correspond to the prisoner's dilemma), finite value of $K$ can avert TOC if there are mostly cooperators initially. 

When $\sf U_0$ corresponds to the coordination game, the extent of aversion of the TOC depends on the relative strength of $-\delta_0$ and $\theta$. As can be seen above the oblique line ($-\delta_0=\theta$) in FIG.~\ref{ToC}(a),  $-\delta_0>\theta$; it means that the enhancement rate by the cooperators is overshadowed by the fact that a player has more affinity to defect against a cooperator than cooperate against a  defector when compared with the region given by  $-\delta_0<\theta$ below the oblique line. Naturally, the extent of the TOC should be more in the former. Similar consideration explains the difference in the extent of the TOC below and above the line, $-\delta_0=\theta$, that splits the region of the anti-coordination game as well.

In the case of the harmony game, if $K$ is made finite, the TOC results when initially there is dearth of cooperators. The difference in the nature of $x_0$-$n_0$ plot near $n_0=0$ above and below the line may be understood physically as follows: Above the line, $\delta_0>\delta_1$ and below the line $\delta_0<\delta_1$. In other words, in the former there is comparatively more affinity to defect against a cooperator than against a defector. Hence, if initially there are a small number of cooperators, in the former case the defectors simple overcome them which is harder to do in the latter case because the affinity to defect against the cooperator is less.

\section{The mechanism}
\label{sec:mech}

\textcolor{black}{ Having discussed some important aspects of the results from physical considerations, we now bring a crucial physical feature of the system to the fore. It is the phenomenon of bistability in the presence of finite carrying capacity. The bistability is an omnipresent feature of many physical, chemical, and biological systems; there is a lot of current interest in its presence in the phenomena witnessed in quantum~\cite{Roberts2020,Landa2020,Yang2021}, thermal~\cite{Nguyen2020,Wang2020,Paul2021}, electrical~\cite{Tadokoro2021}, optical~\cite{Parmee2021,Schmidt2021,Parmee2021}, and mechanical~\cite{Alexandrov2020,Kidambi2020, Kozyreff2021,Gayout2021} systems. Consequently, it is exciting to find the bistability in an eco-evolutionary dynamical scenario and that too resulting in counterintuitive conclusions: e.g., the bistability makes it possible to sustain cooperation---and hence to avert the TOC---in the finite population even when every individual faces the prisoner's dilemma.}

\textcolor{black}{ In the context of the eco-evolutionary dynamics, the bistability manifests itself as the partition of the set of all possible initial conditions into two classes---one class leads to the complete realization of the TOC, whereas the other leads to the complete aversion of the TOC. In this section, we intend to explain the mathematical mechanism leading to the related observed effects of the finite carrying capacity. {A closely related question} is how the results for finite carrying capacity tend towards the results known for an infinite population's eco-evolutionary dynamics. }

\textcolor{black}{ We note in FIG.~\ref{ToC}, the results obtained for $K\rightarrow\infty$ are, indeed,  in complete agreement with similar results in literature~\cite{weitz2016oscillating} where the population is supposed to be fixed and infinite. Despite the agreement, we must mention here the important difference with reference~\cite{weitz2016oscillating} which tacitly assumes that the population is fixed and infinite. There the state marked by $(x^*,n^*)=(1,1)$, corresponding to the TOC aversion, is an unstable fixed point and hence there is no possibility of bistability. This is quite unlike the situation of the growing population with infinite carrying capacity where the bistability effectively vanishes not because $(x^*,n^*)=(1,1)$ is unstable but because its basin of attraction vanishes that otherwise is present for any finite carrying capacity. We elaborate on this in what follows.}
		\begin{figure}
		\centering
		\includegraphics[width=1.0\linewidth]{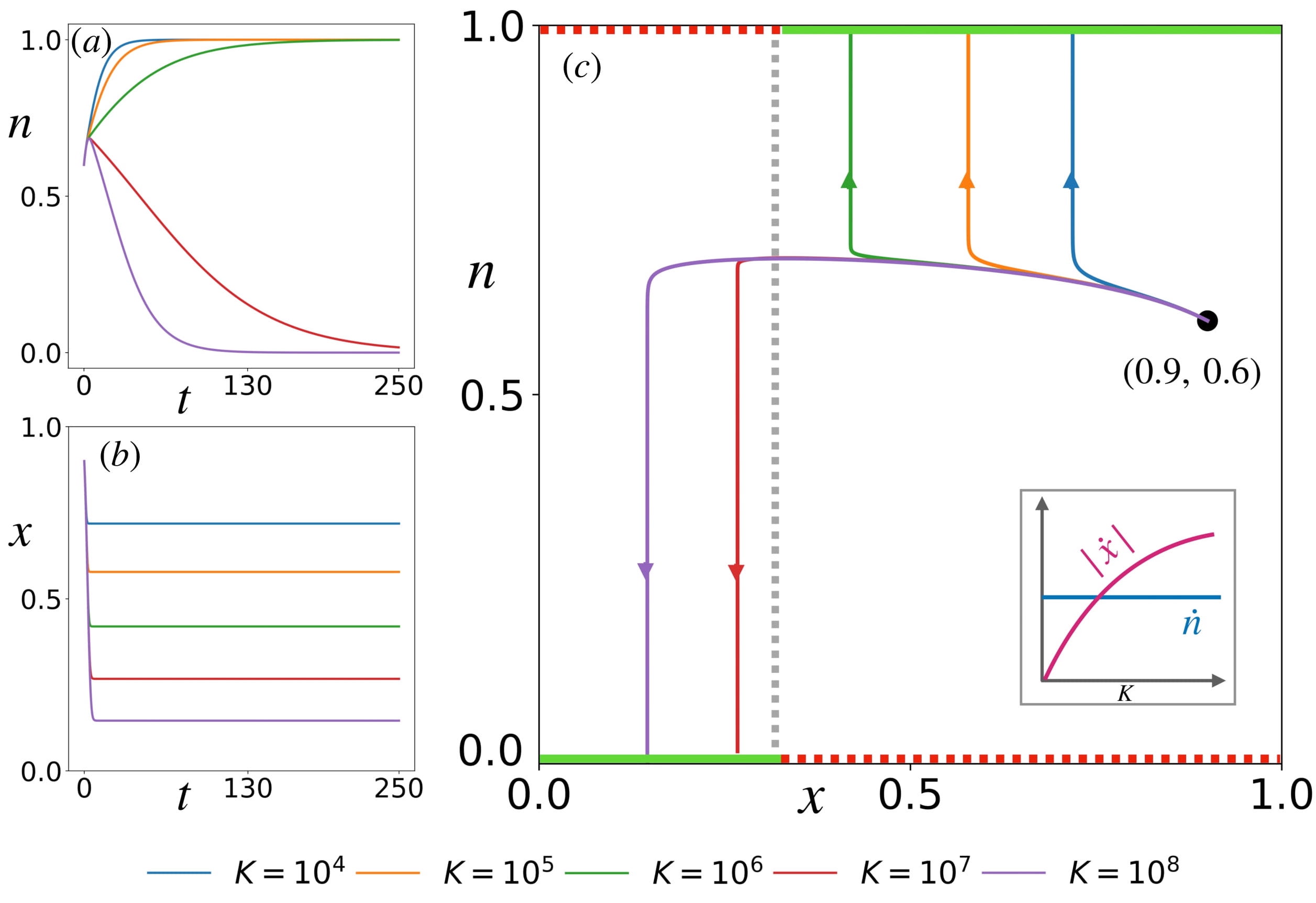}		
		\caption{\textcolor{black}{ Bistability in the growing population with finite carrying capacity: The case of ${\sf U_0}$ for the harmony game with $\delta_0<\delta_1$ is considered here. For the sake of generating illustrative plots, we have specifically chosen ${\sf U_0}=
\tiny{\begin{bmatrix} 
3.5 & 1 \\
2 & 0.05\\
\end{bmatrix}}
$ and ${\sf U_1}=
\tiny{\begin{bmatrix} 
4 & 1 \\
7 & 2\\
\end{bmatrix}}
$. We have fixed $N_0=10^2$ (other values give similar results). The figure exhibits how an initial condition, $(x_0,n_0)=(0.5,0.1)$ approaches either $n^*=0$ or $n^*=1$ with monotonic change in the value of the carrying capacity, $K$. In each subplot, (a), (b), and (c) respectively depict the time-series of $n$, the time-series of $x$, and the phase trajectory projected on $x$-$n$ plane. The solid green line means stable fixed points; the dashed red line indicates unstable fixed points; and the grey dashed line [$x=1/(1+\theta)=2/5$] shows the fixed points for which the linear stability analysis fails. The inset showcases that any increase carrying capacity reduces the time scale of the dynamics of $x$ but has no explicit effect on the time scale of $n$.}}
		\label{fig:velocity1}
	\end{figure}
\begin{figure}[t]
	\centering
	\includegraphics[width=1.0\linewidth]{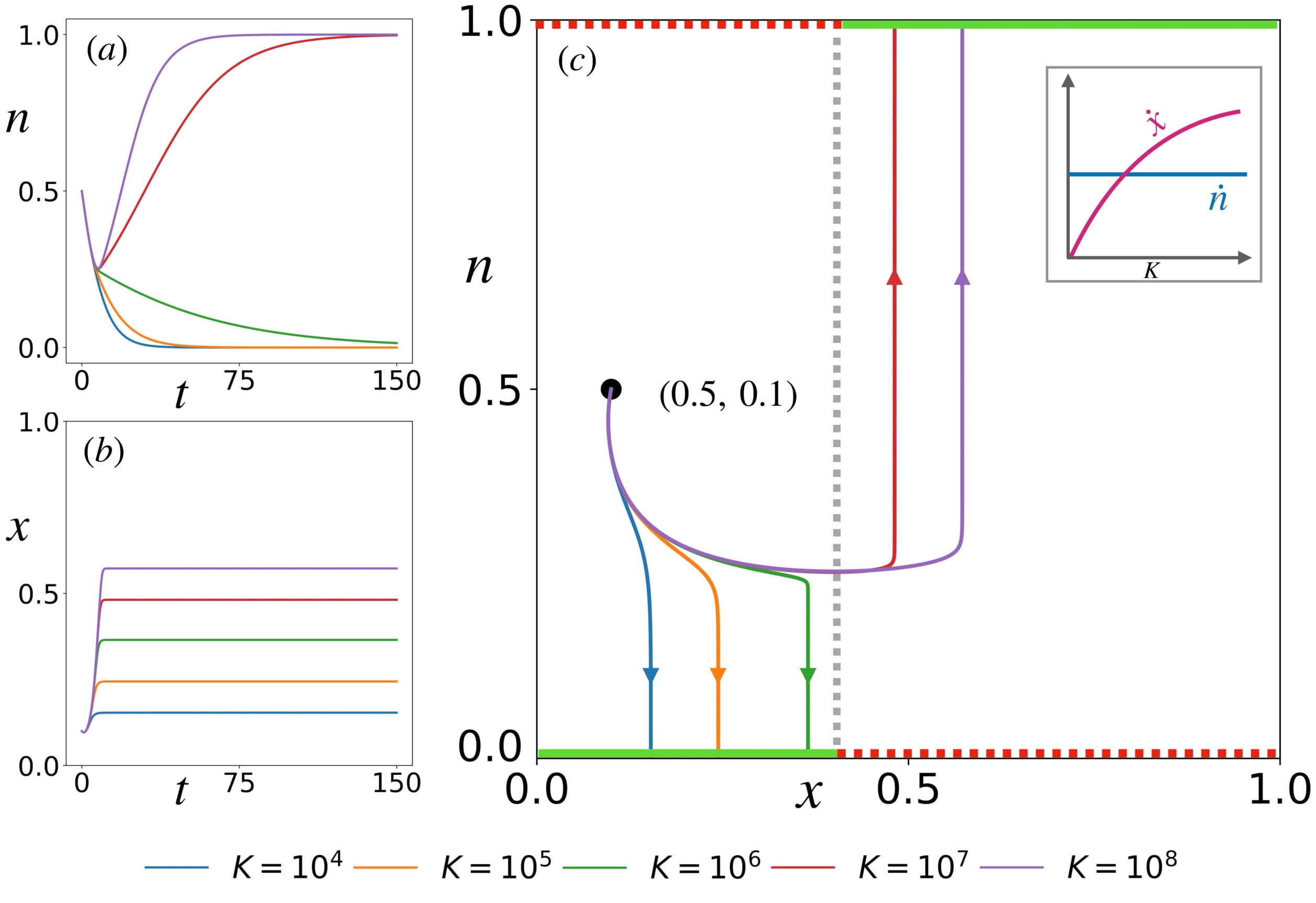}		
	\caption{\textcolor{black}{ Bistability in the growing population with finite carrying capacity: The case of ${\sf U_0}$ for the prisoner's dilemma is considered here. All the details of the plot are same as in FIG.~\ref{fig:velocity1} except that here ${\sf U_0}=
\tiny{\begin{bmatrix} 
2 & 0.05 \\
3.5 & 1\\
\end{bmatrix}}
$, ${\sf U_1}=
\tiny{\begin{bmatrix} 
4 & 1 \\
4.5 & 1.25\\
\end{bmatrix}}
$, $(x_0,n_0)=(0.9,0.6)$, and the grey dashed line is given by $x=1/(1+\theta)=1/3$.}}
	\label{fig:velocity12}
\end{figure}
	
Broadly speaking, there are two cases of interest for us: (i) There is partial or oscillatory aversion of TOC for infinite $K$, but the TOC can show up for certain initial conditions for finite $K$; and (ii) there is TOC for infinite $K$ irrespective of the initial conditions, but it gets averted for certain initial conditions for finite $K$. This hints us that we can best explain the phenomenon by picking an appropriate initial condition, $(x_0,n_0,N_0)$, and find out  towards which attractor---$n=0$ (realization of the TOC) or $n=1$ (aversion of the TOC)---it approaches as $K$ changes. Before we discuss an illustrative example, we mention that the crux of the matter is that the rate of change of $x$ is dependent on $K$ ($\dot{x}$ increases with $K$) but the rate of change of $n$ is $K$-independent. 

Consider an example of case (i): ${\sf U_0}$ is the pay-off matrix for the harmony game with $\delta_0<\delta_1$. The fixed points $(x^*\in[0,1/(1+\theta)),n^*=0,N^*=K)$ (stable) and $(x^*\in(1/(1+\theta),1],n^*=0,N^*=K)$ (unstable) correspond to the TOC while the fixed points $(x^*\in[0,1/(1+\theta)),n^*=1,N^*=K)$ (unstable) and  $(x^*\in(1/(1+\theta),1],n^*=1,N^*=K)$ (stable) correspond to complete aversion of the TOC. Suppose we start from a half-replete resource and a low fraction of cooperators, say we choose $(x_0,n_0)=(0.1,0.5)$; see FIG.~\ref{fig:velocity1}. Additionally, we choose $N_0$ such that $0\ll N_0\ll K$. When $K$ is small, $\dot{x}<\dot{n}$, the initial state is pulled towards a stable fixed point that corresponds to the TOC ($n=0$) while making minimal excursion towards $x$-direction. But if $K$ is increased, $\dot{x}$ increases as well, the state is pushed beyond $x=1/(1+\theta)$ to be subsequently pulled towards the fully replete resource state ($n=1$). Thus, the basin of attraction that leads to complete circumvention of the TOC increases in size as $K$ increases and its measure becomes unity as $K\rightarrow\infty$. The examples of aforementioned case (ii) can be analogously understood as illustrated in FIG.~\ref{fig:velocity12} with the specific example of ${\sf U_0}$ as the pay-off matrix for the prisoner's dilemma.

However, there is an interesting caveat: In the light of FIG.~\ref{ToC} (and also previously reported work~\cite{weitz2016oscillating}), it is clear that when ${\sf U_0}$ corresponds to the harmony game and the anti-coordination game with $-\delta_0<\theta$, $K\rightarrow\infty$ leads to partial (or oscillatory) aversion of the TOC for all initial conditions; but in the light of FIG.~\ref{fig:velocity1}, it appears that  depending on the value of $K$, a particular initial condition asymptotically either leads to $n=0$ or $n=1$. This discrepancy can be traced to the  nature of the set of nonisolated fixed points $(x^*=1/(1+\theta),n^*\in[0,1],N^*=K)$. Unfortunately, the linear stability analysis fails for this fixed point because on linearization about these phase points, the Jacobian of the vector field of the phase space flow has two zero eigenvalues and one negative eigenvalue, viz., $-[(1-n)(R_0+\theta S_0+\theta T_0+\theta^2 P_0)  +n(R_1+\theta S_1 +\theta T_1 +\theta^2 P_1)]/(1+\theta)^2$. The true nature of the stability property of this set of fixed points can be ascertained numerically. 

\begin{figure}[h!]	
	\centering
	\includegraphics[width=1.0\linewidth]{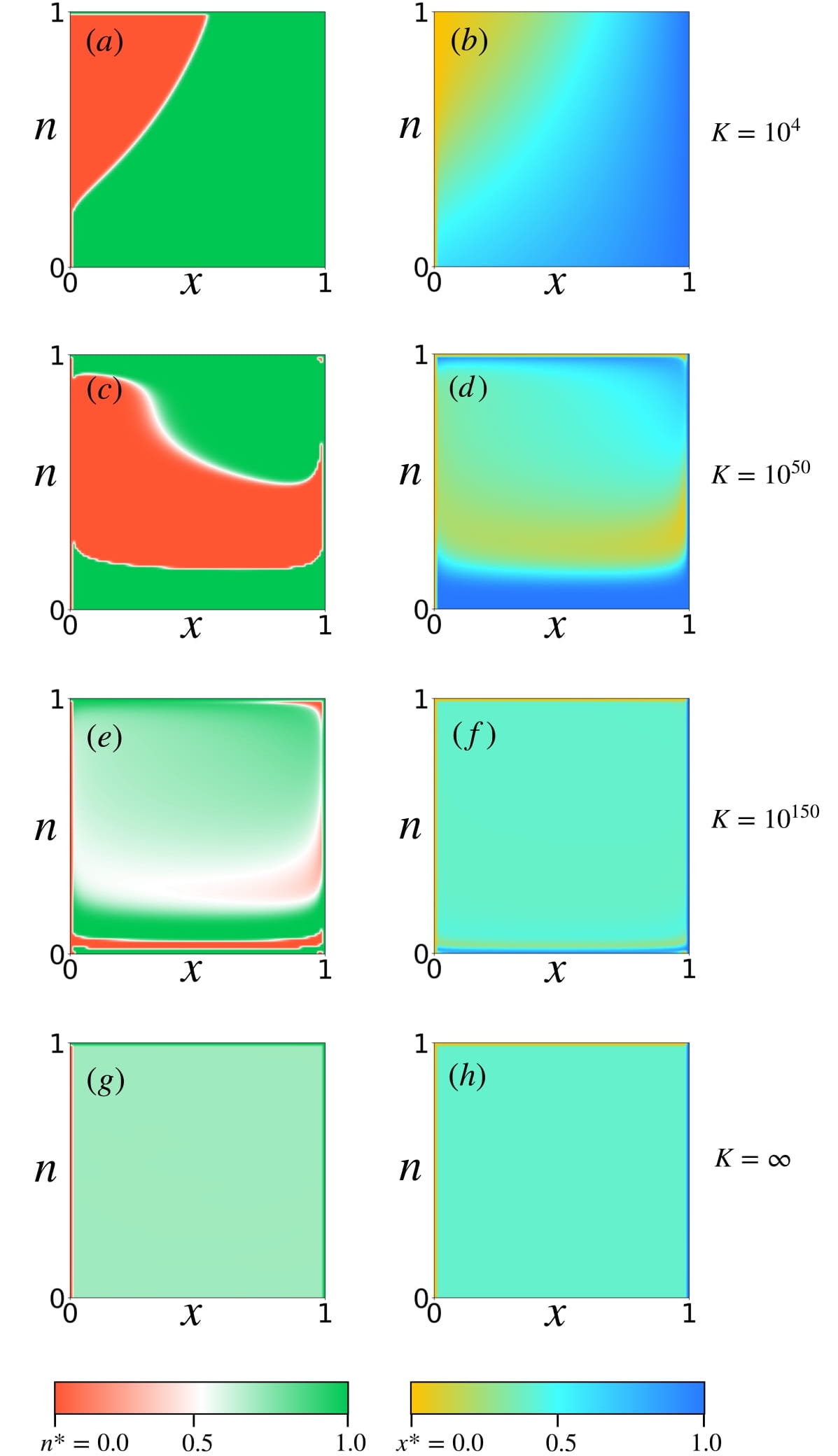}
	\caption{\textcolor{black}{ Exclusive partial aversion of the TOC happens only at $K=\infty$: This figure correspond to ${\sf U_0}$ for the harmony game with $\delta_0<\delta_1$. The first column [subplots (a), (c), (e) and (g)] and the second column [subplots (b), (d), (f) and (h)]  exhibit the value of the final state of the resource $(n^*)$ and the final fraction of the cooperators $(x^*)$ respectively for progressively increasing values of the carrying capacity, $K$, for $100\times100$ uniformly distributed initial conditions $(x_0,n_0)$. We have fixed $N_0=10^2$ (other values give similar results). For  the sake of generating illustrative plots, we have specifically chosen ${\sf U_0}=
\tiny{\begin{bmatrix} 
3.5 & 1 \\
2 & 0.05\\
\end{bmatrix}}
$ and ${\sf U_1}=
\tiny{\begin{bmatrix} 
4 & 1 \\
4.5 & 1.25\\
\end{bmatrix}}
$.}}
	\label{fig:K2infinity}
\end{figure}

In FIG.~\ref{fig:K2infinity}, we explore the stability of the set of nonisolated fixed points, $(x^*=1/(1+\theta),n^*\in[0,1],N^*=K)$, particularly for the same case considered in FIG.~\ref{fig:velocity1}. We find that an interior phase point belonging to this set of fixed points is actually nonlinearly stable and as $K$ increases, its basin of attraction increases in comparison with the other stable fixed points' (corresponding to $n^*=0,1$) basins. In fact, when the carrying capacity becomes infinity, it is the only basin of attraction existing with measure unity. While this observation validates our model that completely reproduces the results existing in the literature~\cite{weitz2016oscillating}), this raises an important point: The partial (or oscillatory) aversion of the TOC seems to be true only in the limit of unrealistically high carrying capacity; what is more plausible is that depending on the initial states of the cooperators and the resource, all three possibilities---complete realization, partial aversion and complete aversion of TOC---may be encountered at any finite carrying capacity howsoever large.
\section{Conclusions}
Our mathematical framework reveals the hitherto unexplored nontrivial effects of the carrying capacity on the TOC in the eco-evolutionary dynamics of game-resource feedback in a growing population. While in the case of a fixed infinite population---realized in our model in the limit of $K\rightarrow\infty$---the status of the TOC is independent of the initial states of the cooperator-fractions and the shared resource, we find that this is definitely not so when the carrying capacity is finite. In fact, depending on the type of strategic game-theoretic interaction and the initial conditions, a finite carrying capacity either averts or causes the TOC; sometimes the result is at direct odds with that would happen, if the carrying capacity is infinite.

Some of the results are rather counter-intuitive: Even though prisoner's dilemma is played, irrespective of whether the common resource is replete or depleted, a finite value of $K$ can avert the TOC if there is enough initial cooperator-fraction. Furthermore, one could argue that finite carrying capacity means smaller number of individuals vying for the shared resource, and hence the aversion of the TOC is intuitively understandable. However, in this context, we have found intriguing scenario where finite $K$ introduces the TOC which is otherwise averted for infinite $K$. This is partially because of the fact that the change in consumption with increasing population ($dn/dN$) increases with decrease in $K$. 

Mathematically, we pin-pointed the mechanism behind the effect of carrying capacity to the fact that the rate of change of cooperator-fraction is dependent on the carrying capacity but the rate of change of the state of the resource is constant with $K$. Our model is validated by the fact that in the limit $K\rightarrow\infty$, we completely reproduce the various scenarios of the TOC as reported in the literature~\cite{weitz2016oscillating}. What, however, is interesting is that the fixed points' stabilities do not change as $K$ approaches infinity, rather it is the basin of attraction of the corresponding stable fixed point that shrinks and practically vanishes making the stable fixed point unattainable.

\textcolor{black}{ It should be pointed out that if one wants to consider the situation of mutually non-exclusive ecological resource and exploitable common resource (as mentioned in Sec.~\ref{sec:introduction}), then mathematically, $K$ should be replaced by the expression $nK_{\text{max}}+(1-n)K_{\text{min}}$ where $K_{\rm max}$ and $K_{\rm min}$ are respectively the maximum and the minimum carrying capacities as the environmental state changes. However, we have found that the qualitatively effects of the finite carrying capacities in this scenario is exactly identical to what we have already reported in this paper; hence, we have chosen not to present the corresponding results to avoid uninformative duplication.}

\textcolor{black}{ The scope of our work, which is based on a deterministic model, is complementary to a very extensive body of published theoretical investigations~\cite{Melbinger2010,Cremer2011,Melbinger2015,Wienand2017,Wienand2018} done using elegant stochastic models. In those models, the changes in the environment have been {captured by a} time-dependent form of the carrying capacity and there is no explicit dynamic equation for the state of the environment. More importantly, {those} models primarily deal with the ecological resources; they do not consider the dynamics of a different shared resource and its exploitation by the replicators in a growing population. We envisage that a stochastic extension of our deterministic formalism, such that the deterministic formalism is seen as its mean-field description, should be the next step of our investigation. Such a stochastic model would relate directly with the aforementioned models if the ecological and the common resource are one and identical.}

We believe that our mathematical framework opens up possibility of further exciting research avenue in the game-resource feedback dynamics in changing population. Specifically, one could consider a renewable resource with intrinsic growth dynamics~\cite{tilman2020nc} and investigate the effect of finite carrying capacity on that. Moreover, one could develop microscopic stochastic birth-death models~\cite{traulsen2005prl,lin2019prl,mendez2015stochastic,2021MCChaos} for both the population and the resource, and investigate the effect of finite population on the eco-evolutionary dynamics in a more fundamental manner. Of course, it is always interesting to extend the results of this paper to the situations of multiplayer~\cite{gokhale2010}, multi-strategy interactions (like in the public goods game~\cite{hofbauer2009geb,Nunn1978,Marwell1979,cornes1996theory}). Additionally, it may also be possible to introduce the phenomenon of cultural evolution~\cite{Laland1995,Feldman1996,Laland2000} because in many cases the TOC is due to cultural practices. 

\textcolor{black}{ Most importantly, however, it needs to be investigated {whether} the effects of the finite carrying capacity predicted in this paper are actually realized in realistic eco-evolutionary systems. In the past, some studies on the interplay between evolution and ecology in growing populations {have} been conducted experimentally, e.g., in yeast population~\cite{sanchez2013feedback} and in bacterial population~\cite{Wienand2015, Becker2018}.  We hope that some experiments with microbes~\cite{Korolev2011_1,Korolev2011_2,Frey2010,Pfeiffer2005,Li2015,Lenski2001} may possibly be designed to this end.}
\acknowledgements
SC is grateful to Archan Mukhopadhyay for insightful discussions. Research of DC has been supported by a J.C. Bose National fellowship (SERB, India).

	\bibliography{Bairagya_etal.bib}
\end{document}